\documentclass[11pt]{article}
\usepackage{a4}

\def\al{\alpha}  
 
\def\ga{\gamma}

\def\bx{{\mathbf{x}}}

\def\br{{\mathbf{r}}}
\def\bk{{\mathbf{k}}}

\newcommand{\ben}{\begin{equation}}
\newcommand{\een}{\end{equation}}
\newcommand{\bea}{\begin{eqnarray}}
\newcommand{\eea}{\end{eqnarray}}
\newcommand{\ba}{\begin{array}}
\newcommand{\ea}{\end{array}}
\newcommand{\bi}{\begin{itemize}}
\newcommand{\ei}{\end{itemize}}

\def\math{\mathsurround 0pt}
\def\oversim#1#2{\lower.5pt\vbox{\baselineskip0pt \lineskip-.5pt
        \ialign{$\math#1\hfil##\hfil$\crcr#2\crcr{\scriptstyle\sim}\crcr}}}

\def\pa{\partial}

\def\na{\nabla}

\def\vev#1{\left\langle #1 \right\rangle}

\title{\textbf{Ordering Dynamics of Topological Defect Networks}
\footnote{Talk given at Solitons 97, July 20--26 1997, Kingston ON, Canada}}
\author{Mark Hindmarsh\\[3pt]
\textit{Centre for Theoretical Physics}\\
\textit{University of Sussex}\\
\textit{Brighton BN1 9QJ}\\
\textit{U.K.}}
\date{\relax}

\begin{document}

\maketitle
\vspace{-80mm}
\begin{flushright}
SUSX-TH-97-024\\
hep-ph/9712490
\end{flushright}
\vspace{80mm}

\begin{abstract}
I report on some work in progress on the dynamics of extended 
objects in field theories after a rapid phase transition, as 
is relevant in the early Universe.  An analytic technique, 
originally introduced to approximate the dynamics of topological defects
in condensed matter systems, is extended to cover relativistic 
objects.  An exact formula for the area density of domain walls in $D$
dimensions is derived.
\end{abstract}

The realisation that spontaneous symmetry breaking in field theories 
might actually be a
dynamic phenomenon in the early Universe brought with it the possibility
that topological defects were formed in the initial stages of the Big
Bang \cite{Kib76}.  The study of the formation and evolution of topological defects
therefore has great implications for particle physics and 
cosmology, as well as being of interest in itself as a difficult problem
in the non-linear dynamics of fields.  The fundamental picture that has
emerged is that networks of topological defects exhibit self-similar
behaviour \cite{VilShe94,HinKib95}.  
The network is described by a length scale $\xi(t)$, which
can be thought of as the average curvature radius of the $p$-dimensional
surface, and this scale increases as a power of time. In the context of
the early Universe, this power seems always to be 1, with a constant of 
proportionality of order 1. Hence the defects
straighten out as fast as causality will allow.

In this talk I report of some work on an analytic approach to the
dynamics of defects formed in a phase transition in the early Universe
\cite{Hin96}.
The programme is as yet in its early stages, but already has produced
interesting results for the dynamics of domain walls, explaining the
results of numerical simulations in a much more detailed way than the
qualitative arguments that went before.  

The analytic approach is based on a 
method which made its first appearance in the condensed 
matter literature \cite{OhtJasKaw82,Bra94}. 
The defects are replaced by scalar fields $u^A(x)$ 
specially designed so their 
zeros track the positions of the defects.  The resulting 
equations of motion are simpler to deal with than the original
field equations, particularly in the 
case of gauge field theories. The initial condition for the field are selected from a 
Gaussian distribution, with mean $U^A(\eta_i)=\vev{u^A(\eta_i,\bx)}$ and 
two-point correlation function $C(\eta_i,\br) = \vev{u^A(\bx)u^B(\bx+\br)\delta
_{AB}/N}$. 
This is supposed to represent an initial high-temperature thermal state.
One can then average the equations of motion for the $u$ fields,
to derive a linearised approximation, which is exactly soluble.
Important quantities such as the defect density can then be found as a 
Gaussian average over the appropriate functions of $u$.
The result is always an expression involving 
powers and derivatives of $C$, or its unequal time version 
$C(\eta,\eta',\bx,\bx') =  \vev{u(\eta,\bx')u(\eta',\bx')}$.

Consider $p$-dimensional defects
in $D$ spatial dimensions (``$p$-branes''). We define with $N=D-p$ real scalar
fields, which vanish precisely at the coordinates of the defects
$X^\mu(\sigma^\alpha)$, where $\mu$ takes the values $0,\ldots,D$ and
$\alpha$ the values $0,\ldots,p$.  Thus we begin with the $N$ equations
\begin{equation}
u^A(X^\mu) = 0.
\end{equation}
Differentiating once with respect to the world-volume coordinates
$\sigma^\alpha$, we find
\begin{equation}
\partial_\beta X^\mu \partial_\mu u^A(X) = 0.
\label{eNorm}
\end{equation}
Thus the $N$ vectors $\partial_\mu u^A(X)$ are spacelike normals 
to the $p$-brane.  

Defects always sweep out surfaces of zero average extrinsic curvature,
and thus the surfaces of constant $u^A$ must also obey this condition. 
The equation that the fields satisfy turns out to be 
\begin{equation}
\Pi^{\mu\nu} (\partial_\mu \partial_\nu - \Gamma^\rho_{\mu\nu} 
\partial_\rho) u^A =0,
\end{equation}
where $\Gamma^\rho_{\mu\nu}$ is the affine connection in the background
space-time, and $\Pi^{\mu\nu}$ is the transverse projector onto the defect
worldsheet. 

At this point we shall specialise to the case $N=1$, domain walls, so that 
the algebra does not obscure the central point. 
The equations of motion may be rewritten in a form which aids the averaging 
procedure:
\begin{equation}
\left[(\partial u)^2 g^{\mu\nu} - \partial^\mu u\partial^\nu u\right] 
(\partial_\mu\partial_\nu u - \Gamma^\rho_{\mu\nu}\partial_\rho 
u) = 0.
\label{eFundWall}
\end{equation}
This equation is linearised
by replacing terms of the third degree by a 
Gaussian two-point correlation function 
multiplied by $u$. For example, 
\ben
\pa^\mu u \pa^\nu u \pa_\rho u \to \vev{\pa^\mu u \pa^\nu u } \pa_\rho u 
+ \mathrm{cyclic.}
\een
We shall suppose that the Universe is described by 
a flat Friedmann-Robertson-Walker space-time, for which the affine 
connection is
$
\Gamma^\rho_{\mu\nu} = (\delta_\mu^\rho\delta_\nu^0 
+ \delta_\nu^\rho\delta_\mu^0 - g_{\mu\nu}g^{\rho 0})(\dot a/a)$. Here, 
$a(\eta)$ is the scale factor and $\eta$ is conformal time. The linearised 
We define the following correlators  
\begin{equation}
T(\eta) = \vev{\dot u(\eta,\bx)\dot u(\eta,\bx)}, \quad 
S(\eta) = \frac{1}{D}\vev{\na u(\eta,\bx)\na u(\eta,\bx)}. 
\end{equation}
and note that $C(\eta) = C(\eta,\mathbf{0})$.
It can be shown that the linearised equations of motion then take the form
\begin{equation}
\ddot u + \frac{\mu(\eta)}{\eta} \dot u - v^2\nabla^2u = 0,
\label{eLEOM}
\end{equation}
where $\mu(\eta)$ and $v$ are functions of $D$.
For Friedmann models, one can show that 
\begin{eqnarray}
\mu(\eta) &= &	-2\eta(\dot S /S) + 
	{\alpha(\eta)}\left[D  - 
	3\left({T}/{S}\right)\right],
\\
v^2 &=& 	\left[{D-1} - \left({T}/{S}\right)\right]/D,
\end{eqnarray}
where $\alpha(\eta) = \eta\dot a/a$.

We expect scaling solutions to have $S,T \propto \eta^\delta$, and 
so as long as we are not near a transition in the equation of state 
of the Universe (such as that between the radiation- and matter-dominated
eras), $\mu$ and $v^2$ are constant. Thus, imposing the boundary 
condition that $u$ be regular as $\eta \to 0$,  
(\ref{eLEOM}) has the simple solution
\begin{equation}
u_{\bk}(\eta)  = A \left(\frac{\eta}{\eta_{\rm i}}\right)^{(1-
\mu)/2+\nu} \frac{J_\nu(kv\eta)}{(kv\eta)^\nu},
\end{equation}
where $(1-\mu)^2/4 = \nu^2$. The form of the initial power spectrum 
$P_{\rm i}(k) = |u_{\bk}(\eta_{\rm i})|^2$ is taken to be 
white noise, behaving as $k^0$ at small $k$, as appropriate to an
uncorrelated field.

We can substitute the solution back into 
the correlation functions $C$, $S$, and $T$.
A self-consistent solution for $\mu$ turns out only to be possible if we 
take $\nu=(\mu-1)/2$, so that $\eta(\dot S/S) = -(D+2)$. 
We may thus obtain  an algebraic equation for $\mu$ 
with $\al = \eta(\dot a/a)$ as a parameter.
For example, in Minkowski space ($\al = 0$) we find $\mu = 2(D+2)$.
We can now 
calculate anything that can be expressed in terms of local functions of 
the field and its derivatives, by averaging over the Gaussian probability 
distribution for the fields.  The easiest quantity to calculate is the comoving 
area density of the walls,
\begin{equation}
{\cal A} = \int d^D\sigma\,\sqrt{-\gamma}\, \delta_{D+1} 
(x-X(\sigma)).
\end{equation}
Making the coordinate transformation from $x^\mu$ to $(\sigma^\alpha,
u)$ near the wall, this can be rewritten as 
\begin{equation}
{\cal A}  = \delta(u)|\partial u|.
\end{equation}

The probability distribution takes the form
\begin{equation}
P(\partial_\mu u, u) = {\mathcal{N} }\exp \left[
-\frac{1}{2} ( u,\partial_\mu u) \left(
\begin{array}{cc}
X & Y^\nu \\
Y^\mu & Z^{\mu\nu}
\end{array}\right)
\left(
\begin{array}{c}
u\\
\partial_\nu u
\end{array}\right)\right],
\end{equation}
where
$$
X =T/SC\ga, \quad
Y^\mu = -\frac12\delta^{\mu}_{0} (1/S\gamma),\quad
Z^{00} = 1/S\gamma, \quad
Z^{ij} = \delta^{ij}(1/S), 
$$
and $\gamma = \left(\frac{T}{S} - 
	\frac{1}{4}\frac{\dot C^2}{CS}\right).
$
The normalisation of the probability distribution is given by
${\mathcal{N}} = (2\pi)^{-5/2}(CTS^3)^{-1/2}$.
We are now in a position to evaluate the average area density.
A long calculation shows that
\begin{equation}
\langle {\mathcal{A} }\rangle = 
\sqrt{\frac{S}{\pi C}} \frac{\Gamma((D+2)/2)}{\Gamma((D+1)/2)}
F\left(\frac12,-\frac12;\frac{D+1}{2};1-\gamma\right).
\end{equation}

Recalling the behaviour of $S$ and $C$ with conformal time, 
we immediately see that we reproduce the correct scaling 
behaviour for the area density, ${\mathcal{A}} \propto \eta^{-1}$.  
But this technique also gives the coefficient of proportionality, which 
we can compare with numerical 
simulations.  
The comparison in $D=3$ is quite satisfying \cite{Hin96}. For example, 
simulations \cite{PreRydSpe89,CouLalOvr95,LarSarWhi96} give 
${\mathcal A} \simeq 1.5\eta^{-1}$ in a radiation dominated Universe, 
while the theoretical prediction is $2.1\eta^{-1}$.
In $D=2$ the agreement is also good, although 
there is evidence for a small deviation from the $\eta^{-1}$ 
behaviour in $D=2$ \cite{PreRydSpe89,CouLalOvr95}.  
However, it should be borne in mind that the theoretical approach 
involves an approximation which is not well-controlled: replacing 
correlators by their Gaussian averages. This is not an expansion in
a small parameter, so assuming that the higher order corrections are
small.

I wish to thank Alan Bray and
Ed Copeland for many useful discussions and help.
The author is supported by PPARC Advanced
Fellowship number B/93/AF/1642, by PPARC grant GR/K55967, and by 
the European Commission under the Human Capital and Mobility
programme, contract no.~CHRX-CT94-0423.


\begin{thebibliography}{99}
\bibitem{Bra94} A. Bray. {\em Adv. Phys.}, D43:357, 1994.
\bibitem{CouLalOvr95} D. Coulson, Z. Lalak, and B. Ovrut.
Phys. Rev., D53:4237, 1996.
\bibitem{Hin96} M. Hindmarsh. {\em Phys. Rev. Lett.}, D77:4495, 1996.
\bibitem{HinKib95} M. Hindmarsh and T.W.B. Kibble. {\em Rep. Prog. Phys.},
55:478 (1995).
\bibitem{Kib76} T.W.B. Kibble. {\em J. Phys.} {\bf A} (1976).
\bibitem{LarSarWhi96} S. Larsson, S. Sarkar and P. White. 
Phys. Rev. D55:5129, 1997.
\bibitem{OhtJasKaw82} T. Ohta, D. Jasnow, and K. Kawasaki. 
{\em Phys. Rev. Lett.} 49:1223, 1982.
\bibitem{PreRydSpe89} W.H. Press, B.S. Ryden, and D.N. Spergel,
{\em Ap. J.} 347:590, 1989.
\bibitem{VilShe94} A. Vilenkin and E.P.S. Shellard. {\em Cosmic Strings
and Other Topological Defects}, Cambridge Univ. Press, Cambridge, 1994.
\end{thebibliography}
\end{document}